\documentclass[useAMS,usenatbib,usegraphicx]{mn2e}
\usepackage{color}
\definecolor{dark-blue}{rgb}{0.15,0.15,0.4}
\usepackage[ocgcolorlinks]{hyperref}
\hypersetup{citecolor=dark-blue, linkcolor=dark-blue}
\usepackage{amsmath}
\usepackage[T1]{fontenc}
\usepackage{aecompl}
\usepackage{amsfonts}
\usepackage{amssymb}
\usepackage{times}

\title[ULX in VII~Zw~403]{Transition of an X-ray binary to the hard ultraluminous state in the blue compact dwarf galaxy VII~Zw~403}
\author[M. Brorby, P. Kaaret, H. Feng]{M. Brorby$^{1}$\thanks{E-mail:
matthew-brorby@uiowa.edu}, P. Kaaret$^{1}$, and H. Feng$^{2}$\\
$^{1}$Department of Physics and Astronomy, University of Iowa, Iowa City, IA 52242\\
$^{2}$Department of Engineering Physics and Center for Astrophysics, Tsinghua University, Beijing 100084, China}
\begin{document}

\pagerange{\pageref{firstpage}--\pageref{lastpage}} \pubyear{2014}

\maketitle

\label{firstpage}

\begin{abstract}
We examine the X-ray spectra of VII Zw 403, a nearby low-metallicity blue compact dwarf (BCD) galaxy. The galaxy has been observed to contain an X-ray source, likely a high mass X-ray binary (HMXB), with a luminosity of $1.3-23\times 10^{38}$~erg~s$^{-1}$ in the $0.3-8$~keV energy range. A new Suzaku observation shows a transition to a luminosity of $1.7\times 10^{40}$~erg~s$^{-1}$ [$0.3-8$~keV], higher by a factor of $7-130$. The spectra from the high flux state are hard, best described by a disk plus Comptonization model, and exhibit curvature at energies above 5~keV. This is consistent with many high-quality ultraluminous X-ray source spectra which have been interpreted as stellar mass black holes (StMBH) accreting at super-Eddington rates. However, this lies in contrast to another HMXB in a low-metallicity BCD, I~Zw~18, that exhibits a soft spectrum at high flux, similar to Galactic black hole binaries and has been interpreted as a possible intermediate mass black hole. Determining the spectral properties of HMXBs in BCDs has important implications for models of the Epoch of Reionization. It is thought that the main component of X-ray heating in the early universe was dominated by HMXBs within the first galaxies. Early galaxies were small, metal-deficient, star forming galaxies with large H~I mass fractions --- properties shared by local BCDs we see today. Understanding the spectral evolution of HMXBs in early universe analogue galaxies, such as BCDs, is an important step in estimating their contribution to the heating of the intergalactic medium during the Epoch of Reionization. The strong contrast between the properties of the only two spectroscopically studied HMXBs within BCDs motivates further study on larger samples of HMXBs in low metallicity environments in order to properly estimate the X-ray heating in the early universe.
\end{abstract}


\section{Introduction}\label{intro}
The first X-ray sources in the early universe were binary systems containing compact objects formed in the collapse of massive stars in low metallicity environments. During the Epoch of Reionization, X-rays from these high mass X-ray binaries (HMXBs) may have contributed to the heating of the intergalactic medium (IGM)~\citep{shull1985, haardt1996, mirabel2011, mcquinn2012, mesinger2013}. 

Recent models \citep[e.g.,][]{venkatesan2001,ricotti2004,linden2010,fragos2013a} and observations \citep[e.g.,][]{mapelli2010,kaaret2011,prestwich2013,basu-zych2013,brorby2014} suggest that X-ray emission and binary populations are enhanced (for a given star formation rate) in regions of low-metallicity. The theoretical studies find that: (1) lower metallicity produces weaker stellar winds, which result in a greater number of more massive black holes, and thus more massive black hole HMXBs~\citep{fragos2013a}; and (2) that the HMXB populations in low metallicity environments are dominated by Roche lobe overflow HMXBs which are very effective in transferring mass to the compact object in comparison to the wind-driven, supergiant HMXBs that dominate in high metallicity environments~\citep{linden2010}. The observational studies find that the X-ray luminosity function describing the population of HMXBs remains relatively flat ($dN/dL \propto L^{-\alpha}$, where $\alpha=1.6$), and therefore the total X-ray luminosity of the star-forming region is dominated by the most luminous sources. The bright end of the X-ray luminosity function consists of the aptly named ultraluminous X-ray sources (ULXs). 

Ultraluminous X-ray sources are bright ($L_X>10^{39}$~erg~s$^{-1}$) extragalactic sources that are not associated with the nucleus of their host galaxy. Studies have concluded that ULXs are likely binary systems containing either a neutron star~\citep{bachetti2014}, an intermediate mass black hole (IMBH)~\citep{colbert1999,kaaret2001,dewangan2006}, or a stellar mass black hole (StMBH) with a luminosity that exceeds its Eddington limit~\citep{poutanen2007,gladstone2009,kawashima2012,sutton2013}. The IMBH ULXs would have spectra similar to the StMBH binary systems within our Galaxy, but with higher luminosities~\citep{colbert1999}. However, if the compact object accretes at super-Eddington rates, then its spectrum will likely differ from the standard sub-Eddington black hole binary (BHB) states. \cite{gladstone2009} provide an observationally defined state for a set of high quality ULX spectra called the \textit{ultraluminous state}. This state is characterized by a soft excess modelled by a cool disc with a power law tail and a break in the spectrum above $\sim 3$~keV. The curvature at higher energies is not seen in the standard states of Galactic BHBs and \cite{gladstone2009} identify this state with super-Eddington accretion flows onto StMBHs.

The shape of the emission spectrum of the first X-ray sources plays a key role in understanding their contribution to heating of the IGM in the early universe and modifies constraints on HMXB population in the first galaxies. The spectral shape affects how X-rays heat the IGM because the mean free path and the fraction of energy deposited as heat versus ionization varies depending on the X-ray's energy \citep{mcquinn2012,mesinger2013,pacucci2014}. The distance the X-rays travel and the energy they deposit in the IGM would affect the morphology of the heating and ionization during the Epoch of Reionization and these effects may be seen in future observations of the redshifted 21-cm hydrogen line (e.g., SKA\footnote{\url{https://www.skatelescope.org/}}).

Another effect of spectral shape is the modification of the K-correction for high redshift objects. If the most luminous X-ray sources in the first galaxies exhibit the same type of spectral curvature seen in ultraluminous X-ray sources (ULXs), as opposed to a straight power law, relatively fewer high energy photons would be redshifted into the soft X-ray band. This would relax the constraint imposed by the observed soft X-ray background allowing for a significant enhancement in the X-ray luminosity of early, star-forming galaxies~\citep{kaaret2014}.

In order to determine the spectral shape of ULXs in the early universe, we must study nearby analogues of the first galaxies. Starburst galaxies, such as blue compact dwarf galaxies (BCDs), have X-ray fluxes that are dominated by the most luminous high-mass X-ray binaries within them~\citep{thuan2004}. BCDs are physically small galaxies with blue optical colors, low-metallicities, large H~I mass fractions, and generally exhibit a recent burst of star formation. These dwarf galaxies provide the best available local laboratories for young galaxies in the early universe~\citep{kunth2000}.

\cite{linden2010} showed that for $Z/Z_\odot < 0.2$ the HMXB population of a star forming region should become dominated by Roche lobe overflow HMXBs. By extension, this would lead to a larger ULX population. \cite{prestwich2013} used observations to show that the ULX population is enhanced at low $Z/Z_\odot$ and the transition takes place in the range $0.06 < Z/Z_\odot < 0.13$ ($7.7 < 12 + \log($O/H$) < 8.0$). The galaxies that were below this range $(Z/Z_\odot \lesssim 0.06)$ were deemed extremely metal poor galaxies (XMPGs). It is this population of XMPGs that we chose to focus on, as the earliest galaxies should be very metal poor.

To date, a total of eight low-metallicity BCDs have been found that contain ULXs~\citep{thuan2014} as defined by having a luminosity $> 10^{39}$~erg~s$^{-1}$ (Eddington luminosity of a 10 M$_\odot$ black hole). Of these, six are XMPGs and only two of these BCDs, VII~Zw~403 ($Z/Z_\odot = 0.019$) and I~Zw~18 ($Z/Z_\odot = 0.062$), have had sufficient counts to allow for a spectral study. Other ULXs have been observed in sub-solar metallicity dwarf galaxies (e.g., Holmberg~II: \citealt{zezas1999,kaaret2004}; Holmberg~IX: \citealt{laparola2001,wang2002,miller2004}), and have metallicities ($Z/Z_\odot = 0.1-0.3$~\citep{egorov2013} for Holmberg~II and $Z/Z_\odot \approx 0.09-0.5$~\citep{moustakas2010,makarova2002} for Holmberg~IX) that may fall within the range for enhanced ULX production \citep{linden2010,prestwich2013}. However, for this study we strictly examine only the lowest metallicity galaxies, VII~Zw~403 and I~Zw~18, as these are the best analogues to the extremely metal poor galaxies in the early universe.

In this paper we focus on VII~Zw~403, a low-metallicity BCD that contains a luminous X-ray binary and is at a distance of 4.34~Mpc~\citep{karachentsev2003}.
A recent \emph{Suzaku} observation of VII~Zw~403 shows the X-ray source to be in a much higher flux state than previous observations and the spectrum is similar to many ULX spectra. In Section~\ref{sect:analysis}, we describe the procedures used to prepare and analyze our dataset. Results of timing and spectral analysis are presented as well. We explain the physical motivation for some of the more complex models and discuss them in the context of previous high-quality spectral studies. Within Section~\ref{sect:discussion}, we contrast the spectral evolution of VII~Zw~403 with that of I~Zw~18, another low-metallicity BCD containing an HMXB that has been observed to transition between low and high flux states. The observed spectral state transitions for I~Zw~18 are more analogous to the Galactic BHBs than the ULX spectra, which is inconsistent with the observed ULX-like spectra of VII~Zw~403. Larger spectral studies and more frequent observations of ULXs within XMPGs will allow for a clearer picture to emerge regarding the spectral states exhibited by HMXBs in the early universe. Throughout the paper all fluxes and luminosities are in the 0.3--8~keV energy range, unless otherwise specificied.

\section{Observations \& Analysis}\label{sect:analysis}
\subsection{Timeline of VII~Zw~403 X-ray Observations}
VII~Zw~403 was observed with Chandra on 2000-01-07 for a total of 10.6~ks and about 300 counts were detected. The data were analyzed by \cite{ott2005} who found a good fit with a hard power law model $(\Gamma=1.75)$ and an absorbed flux of $1.5\times 10^{-13}$~erg~cm$^{-2}$~s$^{-1}$ in the 0.3-8.0~keV band. This flux is lower by a factor of 40 than the high, hard state seen in the Suzaku observation discussed in this paper. 

\begin{figure}
\centering
\includegraphics[width=0.49\textwidth]{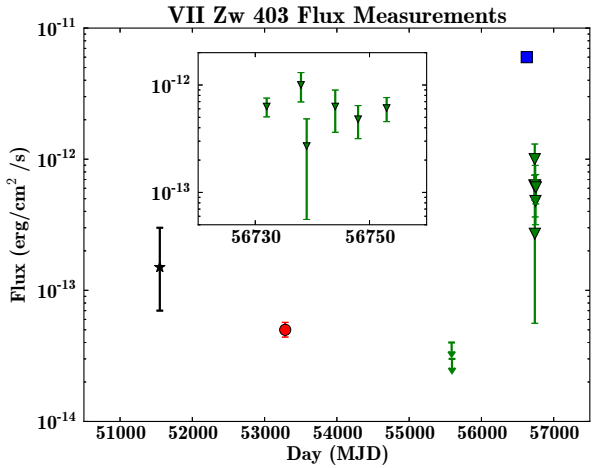}
\caption{Flux measurements of VII~Zw~403 in units of erg~cm$^{-2}$~s$^{-1}$. The triangles (green) are Swift data, star (black) for Chandra, circle (red) for XMM, and square (blue) for Suzaku. This plot is a visual representation of the data in Table~\ref{tab:fluxes}. The inset shows the six Swift observations from 2014. }\label{fig:flux_history}
\end{figure}

There have been a total of 8 observations with \emph{Swift}, two in 2011 and six in 2014. Data from 2011-01-27 and 2011-02-01 show no significant source counts and are limited to fluxes below $4\times 10^{-14}$~cgs. Table~\ref{tab:fluxes} and Figure~\ref{fig:flux_history} show that the data taken from 2014-03-16 to 2014-04-06, roughly 4-5 months after the Suzaku observation, have flux levels varying around an average of $6\times 10^{-13}$~cgs which is 4 times higher than the low state (Chandra) and 10 times lower than the high state (Suzaku). There were not enough counts in the Swift observations to extract individual or summed spectra.

VII~Zw~403 was previously observed with XMM-Newton on 2004-10-07 for 46.9~ks. We created new events files using the \texttt{epproc} and \texttt{emproc} commands with the most recent calibration files. Events were selected from intervals with low
background ($\pm 3 \sigma$ of the mean quiescent level). From this, we found GTIs of 14.9~ks, 33.7~ks, and 32.3~ks for the PN, MOS1, and MOS2 detectors, respectively. Energy spectra were extracted from a circular region with a radius of 20~arcseconds and had a total of 630 net counts from the three detectors. The spectra were rebinned such that each energy bin had at least 15 net counts or was 1/4 of the local FWHM. The fitting was restricted to the 0.3-5.0~keV energy range due to background contamination above 5~keV. We fitted the spectra with a fixed Galactic absorption of $N_H = 3.91\times 10^{20}$~cm$^{-2}$ and a power law. The resulting fit led to a $\chi^2$/D.o.F.~$= 35.5/33$ with a photon index of $\Gamma = 2.1\pm 0.2$ and a model flux of $4.6\times 10^{-14}$~ergs~cm$^{-2}$~s$^{-1}$ (0.3-8.0~keV). This flux is lower than the Suzaku observation by more than a factor of 100.

\begin{table}\large
\caption{Swift observations of VII ZW 403. Count rates and
absorbed fluxes are for the 0.3-8 keV band. Fluxes from Chandra, XMM, and
Suzaku are shown for comparison.}
\centering
\begin{tabular}{lccc}
\hline
\hline\noalign{\vspace{1mm}}
Date		& Exposure	& Net count rate		& Flux \\
			& (s)		& ($10^{-3}$s$^{-1}$)	& ($10^{-13}$~cgs) \\
\hline
2011-01-27	& 4207		& $<0.7$				& $<0.4$ \\
2011-02-01	& 1664		& $<0.6$				& $<0.3$ \\
2014-03-16	& 2851		& 12.6$\pm$2.5			& 6.3 \\
2014-03-22	& 850		& 20.0$\pm$6.1			& 10.0 \\
2014-03-23	& 753		& 5.3$\pm$4.2			& 2.7 \\
2014-03-28	& 801		& 12.5$\pm$5.3			& 6.3 \\
2014-04-01	& 1450		& 9.7$\pm$3.3			& 4.8 \\
2014-04-06	& 1877		& 12.3$\pm$3.1			& 6.1 \\
\hline
2000-01-07	& Chandra	&						& 1.5 \\
2004-10-07	& XMM		& 						& 0.5 \\
2013-12-01	& Suzaku	&						& 60.0 \\
\hline
\end{tabular}
\label{tab:fluxes}
\end{table}

\subsection{Suzaku Observation}\label{subsect:observations}
Our observations of VII~Zw~403 used the three functioning X-ray Imaging Spectrometer (XIS) cameras aboard \emph{Suzaku}, one of which (XIS1) is a back-illuminated (BI) CCD chip and the other two (XIS0,3) are front-illuminated (FI). The observation was made on 2013-12-01 with a total exposure time of 88.66 ks and resulted in 68409 net counts.

The data were reprocessed using the standard HEASARC script \texttt{aepipeline}. Spectra for each of the three detectors were extracted from a circular region centered at $\rmn{RA}(2000)=11^{\rmn{h}}~28^{\rmn{m}}~07\fs6$,
$\rmn{Dec.}~(2000)=+78\degr~59\arcmin~37\farcs 1$ with a radius of 3$^\prime$ (90 per cent encircled energy fraction) and a background region from an annulus with inner and outer radii of 4$^\prime$ and 6$^\prime$, respectively. The spectral data were grouped such that there was a minimum of 40 counts per energy bin.

The spectra were fitted using multiple models in the $0.4-10$~keV energy range. For sufficient sensitivity, we fit the XIS1 (BI) detector data in the $0.4-8.0$~keV energy range and the XIS0,3 data in the $0.6-10.0$~keV range. 

Contamination has been building up on the spacecraft, leading to additional absorption, mainly by carbon. This effect can be seen in comparing identical spectra using two different sets of response files. Response files based on older calibration data (2006) would show more absorption in the models (wabs, tbabs, phabs, etc.) than response files generated with this contamination taken into consideration. For this reason, we generated our own response files for each spectra as opposed to using archived response files from the HEASARC archive. The response files were created by running \texttt{xisrmfgen} and \texttt{xissimarfgen} upon extraction of the spectra.

\subsection{Suzaku Timing Analysis}\label{subsect:timing}
Short term variability (e.g., quasi-periodic oscillations (QPOs), pulsed emission) has been observed in some ULXs \citep[e.g.][]{strohmayer2003,dewangan2006,feng2010}. These features allow for identification of certain types of objects (e.g., pulsars, eclipsing binaries, etc.) and can constrain size and mass of the emitting object in ways the X-ray spectra cannot.

Observations of VII~Zw~403 were taken in the normal timing mode of Suzaku which has a timing resolution of 8~seconds. Light curves were extracted from the same $3^\prime$ circular regions from which the spectra were extracted. Energy filtering was applied such that only photons with energies in the $0.3-8.0$~keV range were accepted. Using the HEASARC tool \texttt{xselect}, we produced light curves with 200s time bins for sufficient counts. The timing analysis that was done without background subtraction showed no significant variability beyond Poisson fluctuations for the XIS0,3 detectors. Analysis of the background-subtracted light curves still showed no significant variation in the XIS0,3 detectors. The XIS0,3 detectors had net count rates of $0.25$ and $0.26$~counts/s, respectively. The sample and expected variances for these detectors were $\sim 1.4\times 10^{-3}$ and are consistent with a constant light curve. 

Assuming the count rate is constant, the XIS1 detector exhibited variances that were much larger than expected from Poisson noise alone as indicated by an F-test of the variances with 1-P(F-test)~$ > 0.99$. The XIS1 detector is more sensitive to background variations than XIS0,3 because it is back-illuminated. The background light curve for XIS1 showed the same strong variation as the source light curve for XIS1 and thus the background variations are the likely cause of the excess variance. To try and remove any effects due to background variation, we subtracted the background counts on a bin by bin basis for the source region. For a given time bin, we found the number of background counts in the background region, scaled this number by the ratio of the areas of the source and background regions, and subtracted this from the total counts in the source region. The errors were added in quadrature with a weighting to account for the difference in extraction region area. This process removed some of the strong variations we were seeing from the XIS1 source region. The variations in the XIS1 detector were reduced but still showed excess variance with a significance of 1-P(F-test)~$ = 0.99$. The net count rate for the XIS1 detector was found to be $0.28$~counts/s with a sample variance of $2.0\times 10^{-3}$ and an expected variance of $1.7\times 10^{-3}$. Due to the noise in the XIS1 detector, we exclude it from further timing analysis.

Power spectra were calculated for each of the FI detector light curves using the method outlined by \cite{leahy1983}. Due to Suzaku's orbit, there are periodic gaps in the data that occur as a result of either Earth occultation or passage through the South Atlantic Anomaly (SAA). These periodic gaps produce large peaks in the FFT of the light curves. To avoid this problem, we calculate power spectra for each continuous set of light curve data and combine the power spectra in the way described by \cite{leahy1983}. For each detector, 27 power spectra were averaged over a frequency range of $0.37 - 62.1$~mHz ($T=2720$~s) using 169 frequency bins. The power in the averaged power spectrum is distributed as a $\chi^2$ random variable with 2N degrees of freedom, where N is the number of power spectra that were averaged together. We found a maximum power of $\sim 3.0$ (Leahy) between the two detectors. The chance probability of detecting this maximum power in any frequency bin is $p=0.76$ and is therefore not significant.

\begin{figure*}
\centering
\includegraphics[width=0.49\textwidth]{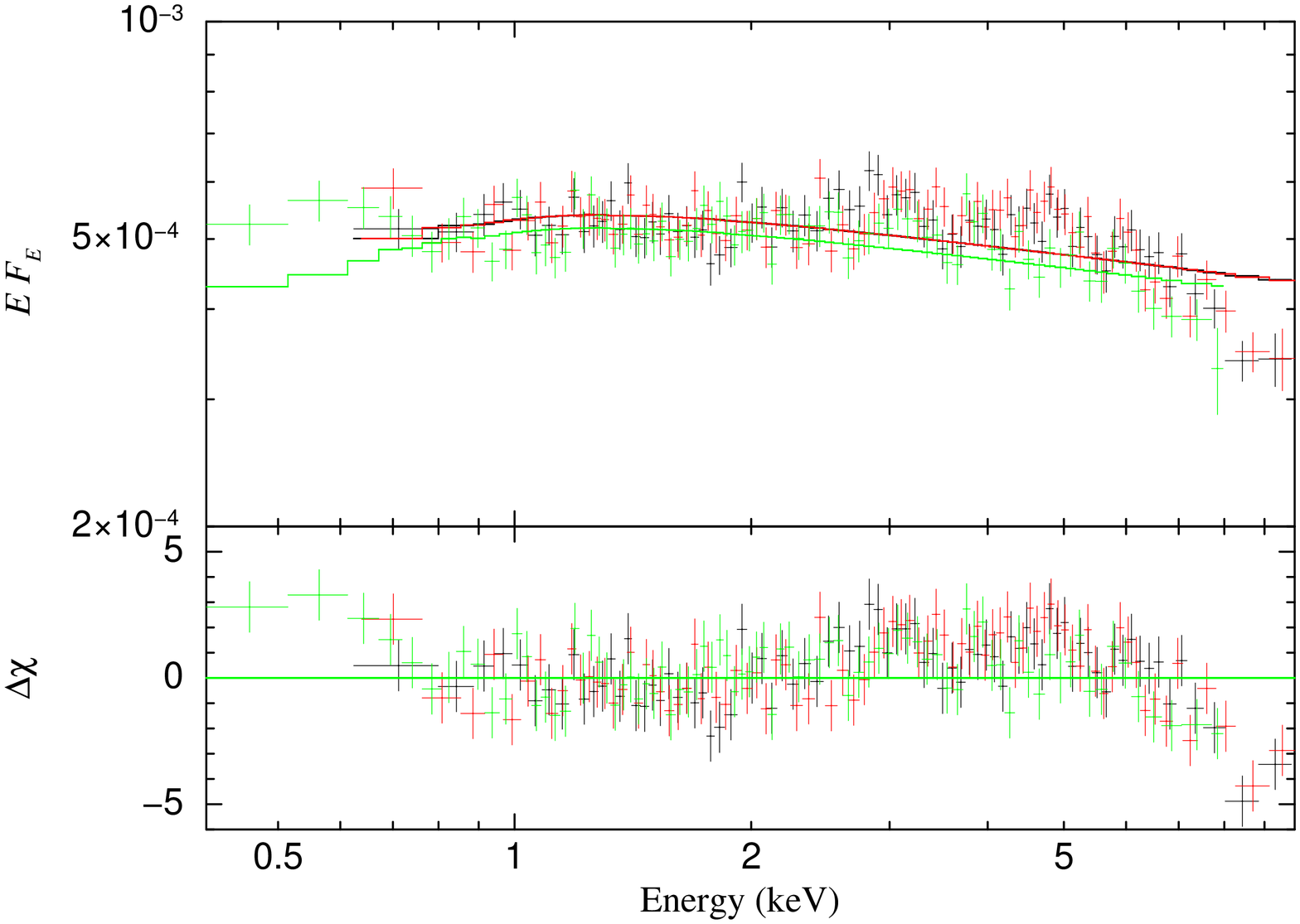}\hfil\includegraphics[width=0.49\textwidth]{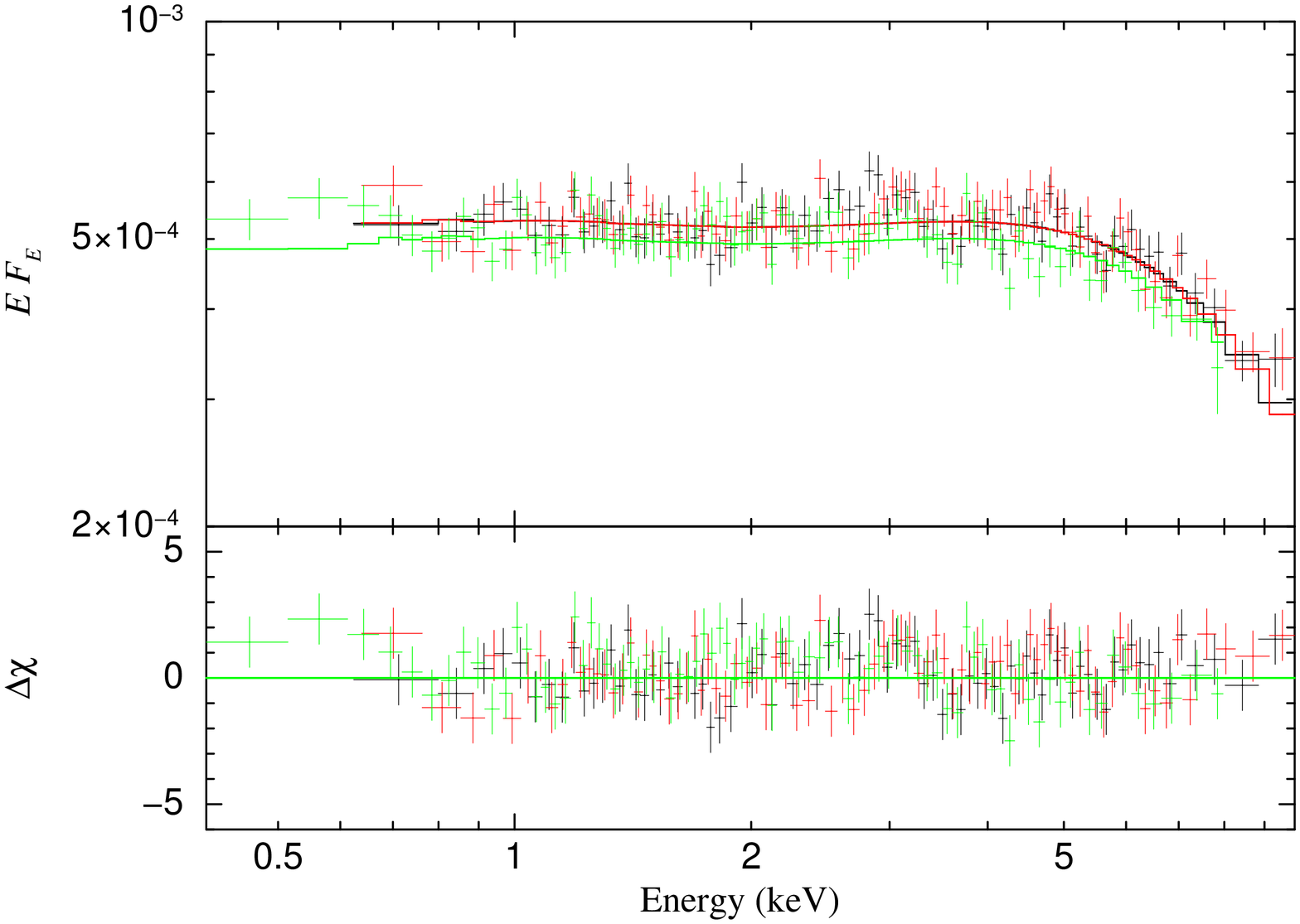} \caption{\textit{Single Component Models}: Suzaku spectra of VII~Zw~403 (XIS0: black, XIS1: green, XIS3: red) rebinned to a minimum of $15\sigma$ or a maximum of 15 channels for visualization. \textit{Left:} Model~(2) power law fit (see Table~\ref{tab:viizw403_final}). \textit{Right:} Comptonization (\texttt{compTT}). The power law models clearly show a break in the spectra above 5~keV and the need for a soft excess at lower energies $(E < 0.8$~keV$)$. The Comptonization model provides an acceptable fit, but is significantly improved by the addition of a soft disk component (see Figure~\ref{fig:diskpn_comptt}).}\label{fig:singles}
\end{figure*}
\subsection{Suzaku Spectral Analysis}\label{subsect:results}
Due to uncertainty in the normalization between the BI and FI detectors, we allowed for a constant factor between the XIS1 and XIS0,3 spectra. This constant factor increases the overall normalization of the XIS0,3 spectra by about 5 per cent.

We began by fitting only single-component models that included interstellar absorption. We define an acceptable fit as having a $\chi^2$ value that lies in the lower 95 per cent of the respective $\chi^2$-distribution. As summarized in Table~\ref{tab:viizw403_final}, the single multicolored disk (\texttt{DISKBB} in \texttt{XSPEC}), simple power law models (\texttt{POWERLAW} in \texttt{XSPEC}), and the exponentially cutoff power law (\texttt{CUTOFF} in \texttt{XSPEC}) did not fit the data well. A broken power law or Comptonization model described the data within acceptable $\chi^2$ values. The broken and cutoff power law models both provided significantly better fits than a single power law model. This is an indication of spectral curvature, a ubiquitous feature in high quality ULX spectra.

The Comptonization model provides the best fit of the single-component models and is the only model that can provide a direct, physical interpretation of the data. It exhibits an extremely cool input photon temperature of $T_0 <0.09$~keV with almost no intrinsic absorption and an electron coronal temperature of $2.15\pm 0.06$~keV with a coronal optical depth of $11.8\pm 0.2$. These parameters are consistent with other high-quality ULX spectra~\citep{gladstone2009}.
By looking at the residuals for these different fits, one can see that the vast improvement is due to the modelling of a clear break in the spectra around 5~keV (see Figure~\ref{fig:singles}). The residuals also indicate that below about $0.8$~keV there is a soft excess that cannot be fit by a single component. This provides the evidence for the second main feature of ULX spectra, a soft excess, and motivates fitting using more complex models.

The appearance of a soft excess in the residuals of our single-component model fits motivates the addition of a second component to the models. The spectra we analyzed have sufficient net counts $(>10~000)$ such that multi-component models may be justified. We fit the spectra with a disk plus powerlaw model (\texttt{diskbb+powerlaw}), and we find two minima in the $\chi^2$ fits (Models 7 and 8). For models (7), (8), and (9) it is assumed that the disk is providing seed photons to the mechanism driving the power law. The fit for which the power law models the soft component and the disk models the hard component (Model 7) has been discussed in other papers~\citep{foschini2004,roberts2005,roberts2007} and was shown to be unphysical. For this reason, the diskbb+powerlaw model (Model 7 in Table~\ref{tab:viizw403_final}) is not physically realizable. We also find that Model (8) does not provide an adequate fit due to the curvature above 5~keV and is improved by the use of an exponentially cutoff power law (Model 9).

\cite{gladstone2009,gladstone2011} provide an observationally defined state for a set of high quality ULX spectra called the \textit{ultraluminous state}. This state is characterized by a soft excess modelled by a cool disc with a power law tail and a break in the spectrum above $\sim 3$~keV, similar to the break observed with Suzaku from VII~Zw~403. In order to compare our spectra with the ULX sample of \cite{gladstone2009}, we use the \texttt{diskpn} model for the disk component as opposed to \texttt{diskbb}. The two disk models are consistent with each other to within 5 percent. Referring to Table~\ref{tab:viizw403_final}, one finds that most of the two-component models provide adequate fits, but only the disk plus Comptonization model provides a physically realistic description of the spectra. In this model, the disk temperature is tied to the input soft photon temperature of the Comptonization model. 
The parameter values for this model agree with the compTT model without the disk component. However, the two-component model provides a fit that is significantly better (1-P(F-test)~$>0.99$) than the compTT model on its own. This is due to improved modelling of the soft excess by the additional disk component.

From the Comptonization models, we find that the fits describe a cool, optically thick corona ($kT_e = 2.2$~keV and $\tau = 11.4$) consistent with values found by \cite{gladstone2009} (i.e., $kT_e \sim 1-3$~keV and $\tau \sim 6-80$). The very cool disc temperature, $T_0 < 0.1$~keV, is consistent with disc temperatures predicted for intermediate mass black holes ($T_0 < 0.5$~keV). However, this is unlikely to be representative of the actual temperature as the presence of an optically thick corona would lead to obscuration of the accretion disc~\citep{kubota2004}, making direct observation of the inner accretion disc impossible. Any physical interpretation based on the inner accretion disc temperature parameter would be questionable and we simply use this parameter as a means to compare to the results of \cite{gladstone2009}.

\cite{sutton2013} divided the ultraluminous state as defined in \cite{gladstone2009} into three spectral subtypes: broadened disc, hard ultraluminous, and soft ultraluminous. The broadened disc state represents a state in which the accretion rate is near the Eddington accretion rate. The hard and soft ultraluminous states are thought to be super-Eddington accretion states. As observed with Suzaku, VII~Zw~403 has a $kT_\text{in} <0.5$~keV and $\Gamma < 2$ (\texttt{diskbb+powerlaw}, Model~8 in Table~\ref{tab:viizw403_final}) which puts it in the hard ultraluminous state. Using the criteria of \cite{sutton2013}, we find the spectral hardness to be $F(1-10$~keV$)/F(0.3-1.0$~keV$)=3.80$, which is the ratio of the unabsorbed fluxes in the indicated energy ranges. This places VII~Zw~403 within the same region as the other hard ultraluminous objects in the hardness--luminosity diagram of \cite{sutton2013}.

\begin{table*}\footnotesize
\begin{minipage}{156mm}
\caption{VII~ZW~403 spectral fitting results: Flux (absorbed~$F_X^\text{abs}$ and unabsorbed~$F_X$), luminosity and fitting are all in the $0.4-10.0$ keV energy band. An unabsorbed flux of $7.0\times 10^{-12}$~cgs converts to a luminosity of $1.6\times 10^{40}$~erg~s$^{-1}$, with an assumed distance of 4.34~Mpc. All models use the \texttt{tbabs} absorption model with Galactic absorption $N_H = 3.91\times 10^{20}$~cm$^{-2}$, which was found using the \texttt{Colden} tool~(\url{http://cxc.harvard.edu/toolkit/colden.jsp}), and the \texttt{tbvarabs} model for the intrinsic absorption such that $Z = 0.062 Z_\odot$.}
\centering
\begin{tabular}{lcccccc}
\hline
\hline\noalign{\vspace{1mm}}
Model	& $\chi^2$/DoF	& $F_X^\text{abs}$	& $F_X$	& $N_H$	& $kT/E_c~^{(a)}$	& $\Gamma$/$\tau$~$^{(b)}$	\\
\hline\noalign{\vspace{1mm}}
		&				& $(10^{-12}$ cgs)	& $(10^{-12}$ cgs)	& $(10^{21}$ cm$^{-2})$ & (keV) & \\
\hline\noalign{\vspace{1mm}}
(1)~Diskbb			& 1749.8/1256	& 6.7	& 6.8	& $<0.009$			& 2.99$\pm$0.06	& $-$\\
(2)~Power law		& 1460.2/1256	& 7.2	& 7.4	& $<0.06$		& 	$-$			& 1.13$\pm$0.01 \\
(3)~Cutoff PL		& 1381.6/1255	& 7.0	& 7.2	& $<0.021$			&17$\pm$3	& 0.95$\pm$0.04 \\
(4)~Broken PL		& 1305.6/1254	& 6.9	& 7.1	& $<0.025$			&5.3$^{+0.7}_{-0.3}$	& 1.07$\pm$0.02, 1.7$^{+0.3}_{-0.1}$ \\
\hline\noalign{\vspace{-2mm}}
\end{tabular}
\begin{tabular}{lccccccc}
\hline\noalign{\vspace{1mm}}
Model	& $\chi^2$/DoF	& $F_X^\text{abs}~^{(c)}$	& $F_X~^{(c)}$	& $N_H$	& $T_0~^{(d)}$	& $kT_e / E_c~^{(e)}$	& $\Gamma / \tau$~$^{(b)}$	\\
\hline\noalign{\vspace{1mm}}
		&				& $(10^{-12}$ cgs)	& $(10^{-12}$ cgs)	& $(10^{21}$ cm$^{-2})$ & (keV) & (keV) & \\
\hline\noalign{\vspace{1mm}}
(5)~CompTT			& 1263.8/1254	& 6.8	& 7.0	& $<0.1$			&$<$0.09 & 2.15$\pm$0.06	& 11.8$\pm$0.2 \\
(6)~Diskpn+compTT	& 1251.1/1253	& 6.9	& 7.5	& 1.4$\pm$0.9	& 0.09$\pm$0.01	& 2.19$\pm$0.08	& 11.4$\pm$0.4 \\
				&			& 0.1 / 6.8	& 0.4 / 7.1		&				&				&				& \\
(7)~Diskbb+PL		& 1250.8/1254	& 6.9	& 7.2	& $<1.1$		&3.6$^{+0.2}_{-0.1}$ & $-$	& 3.0$\pm$ 0.5 \\
				&			& 6.4 / 0.5	& 6.6 / 0.6	&				&		&		&\\
(8)~Diskbb+PL		& 1404.2/1254	& 7.2	& 8.4	& $2.7\pm 0.7$	&0.09$^{+0.01}_{-0.01}$ & $-$	& 1.17$\pm$ 0.02 \\
				&			& 0.2 / 7.0	& 1.0 / 7.4	&				&		&		&\\
(9)~Diskbb+Cutoff	& 1258.9/1253	& 6.9	& 7.1	& $<0.14$		& 0.24$\pm$0.04 & 6.4$^{+1.1}_{-1.0}$	& 0.55$\pm$0.1\\ 
				&			& 0.3 / 6.6	& 0.3 / 6.8	&				&		&		&\\
\hline
\end{tabular}
\label{tab:viizw403_final}
\\
\raggedright
$^{(a)}$ $kT$ is the inner disk temperature for disk models. $E_c$ is the break/cutoff energy for power law models.\\
$^{(b)}$ $\Gamma$ is the photon index for power law models. For compTT, $\tau$ is the plasma optical depth.\\
$^{(c)}$ For the two-component models, individual flux contributions are given below the total flux values using the convention:\\ (disk component flux) / (PL or Comp. flux).\\
$^{(d)}$ $T_0$ is the inner disk temperature.\\
$^{(e)}$ $kT_e$ is the plasma temperature for the compTT models and $E_c$ is the cutoff energy for the exponentially cutoff power law models.

\end{minipage}
\end{table*}
\begin{figure}
\centering
\includegraphics[width=0.45\textwidth]{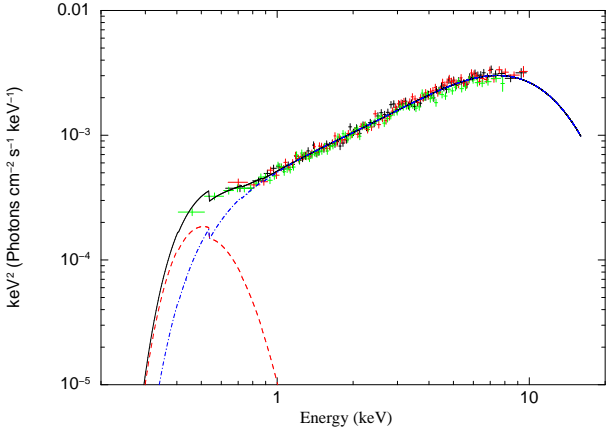}
\caption{Suzaku spectra of VII~Zw~403 (XIS0: black, XIS1: green, XIS3: red) rebinned to a minimum of $15\sigma$ or a maximum of 15 channels for visualization. The model (black line) consists of an absorbed accretion disk (red, dashed line) and a Comptonization component (blue, dash-dotted line). The X-ray spectrum of VII~Zw~403 is quite flat and curvature can be seen at energies $>5$ keV. These spectral properties, along with the observed luminosity of $1.7\times 10^{40}$~erg~s$^{-1}$, are consistent with ultraluminous states found in other extragalactic sources~\citep{gladstone2009}.}\label{fig:diskpn_comptt}
\end{figure}

\section{Discussion \& Conclusions}\label{sect:discussion}
\subsection{Spectral evolution of VII~Zw~403}
Observations of VII~Zw~403 have shown luminosities in the $1.3-23\times 10^{38}$~erg~s$^{-1}$ range $[0.3-8~\text{keV}]$. Table~\ref{tab:fluxes} and Figure~\ref{fig:flux_history} show the known X-ray flux history of VII~Zw~403. While in the low flux state, both Chandra and XMM-Newton observed the X-ray source to have a spectrum consistent with a hard power law ($\Gamma = 1.75-2.1$). Our analysis of a recent Suzaku observation shows that the X-ray source transitioned to a flux more than 100~times brighter ($1.7\times 10^{40}$~erg~s$^{-1}$) than the XMM observation. More recent \textit{Swift} observations indicate that the X-ray source has transitioned out of the high flux state.

The high quality of the Suzaku spectra allowed for the identification of spectral curvature above an energy of $5.3^{+0.7}_{-0.3}$~keV using a broken power~law fit (Model~4 in Table~\ref{tab:viizw403_final}), a feature seen in many high quality ULX spectra. Using a disk plus Comptonization model, we fit the spectra and found our parameters to be consistent with the sample of ULXs in the ultraluminous state studied by \cite{gladstone2009}. This is consistent with the idea that a compact object with an accretion disk is producing photons that are being upscattered by an optically thick corona.

\cite{motch2014} have recently reported that the X-ray source P13, a ULX in NGC 7793, exhibits all the canonical properties of ultraluminous sources and is in a binary system for which optical modulations constrain the black hole mass to be less than 15 solar masses. This X-ray source has also been observed to vary in its X-ray luminosity by at least a factor of 40. Their fitting of a disk plus Comptonization model results in a coronal electron plasma temperature of $kT_e = 2.1^{+0.4}_{-0.2}$~keV and an optical depth of $\tau = 11.7^{+1.2}_{-1.2}$, which are consistent with our results for VII~Zw~403. Other studies \citep[e.g.][]{poutanen2007,gladstone2009,feng2011,kawashima2012,sutton2013}  also support the interpretation that similar ultraluminous sources are StMBH binaries and not the more massive IMBH binaries. Drawing conclusions about VII~Zw~403 from the results of \cite{motch2014} and others would be speculative but a simple comparison may suggest that VII~Zw~403 is a stellar mass black hole undergoing supercritical accretion. More uniform monitoring over a longer interval is needed to adequately constrain the luminosity evolution of the source.

\subsection{Alternative possibilities}
Within VII~Zw~403 there is a second known X-ray source (X-2) that lies within the beam of Suzaku. This source has been observed to have a luminosity 25 times dimmer than the main source (X-1)~\citep{brorby2014} we have discussed in this paper. These two objects are separated by a projected distance of 650~pc or about 30~arcsec. {The relative offsets of the Suzaku source with respect to the well-defined Chandra sources are $20\farcs 8$ for X-1 and $13\farcs 2$ for X-2. Suzaku's absolute pointing has an uncertainty of about $20\farcs$~(Uchiyama 2008). 
The PSF of the XRTs has a broad half-power diameter (HPD) of $\sim 2\arcmin$, but a sharp Gaussian core of width $\sim 10\farcs$. 
Given these two independent errors, the centroid position of the Suzaku source is consistent with both X-1 and X-2.} It is therefore possible that the dimmer source has brightened, but this would require a factor of 1000 change in luminosity, if due solely to this second source. The XMM and Chandra observations are able to resolve these two sources and we find that the flux ratio between X-1 and X-2 decreases by a factor of about 2.7 over the two observations which is consistent with the factor of $3^{+3}_{-1.6}$ change in the flux of X-1 alone (see Table~\ref{tab:fluxes}). Thus, X-1 seems to be the source that is varying, whereas no significant variability is measured for X-2.


From the study presented in this paper, there is no evidence that the main source in VII~Zw~403 is a background AGN. The number of expected background sources within the $D_{25}$ ellipse coincident with VII~Zw~403 is $<0.02$, as determined by the $\log{N}-\log{S}$ curves of \cite{georgakakis2008}. Thus, it is unlikely that either of the two sources associated with VII~Zw~403 are background AGN.

\subsection{Comparison to I~Zw~18}
\cite{kaaret2013} reported a transition from low to high flux for an X-ray binary in the lowest known metallicity BCD, I~Zw~18. The X-ray source was observed to transition from a low-flux state with a hard power law spectrum to a high-flux state with a thermal or exponentially cutoff spectrum. Figure~3 of \cite{kaaret2013} shows that at high-flux the source exhibits curvature over the entire $0.3-10.0$~keV range. This type of high-flux state is consistent with Galactic black hole binaries and lies in stark contrast to the high-flux state of VII~Zw~403. As one can see from Figure~\ref{fig:diskpn_comptt}, VII~Zw~403 shows a relatively flat spectrum and curvature only above 5~keV, which is more consistent with high-quality ULX spectra.

\begin{figure*}
\centering
\includegraphics[width=0.46\textwidth]{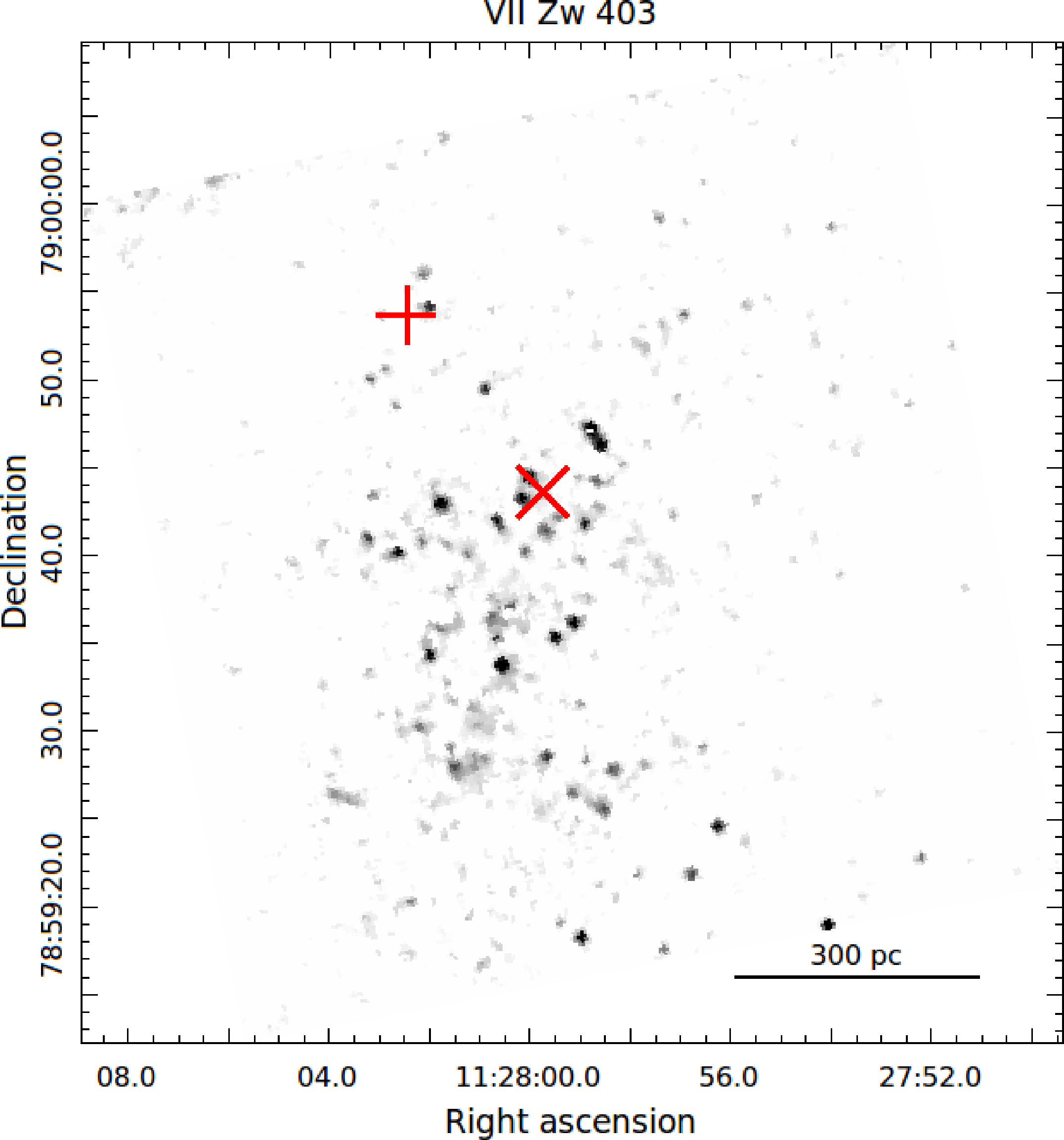}\hfil\includegraphics[width=0.47\textwidth]{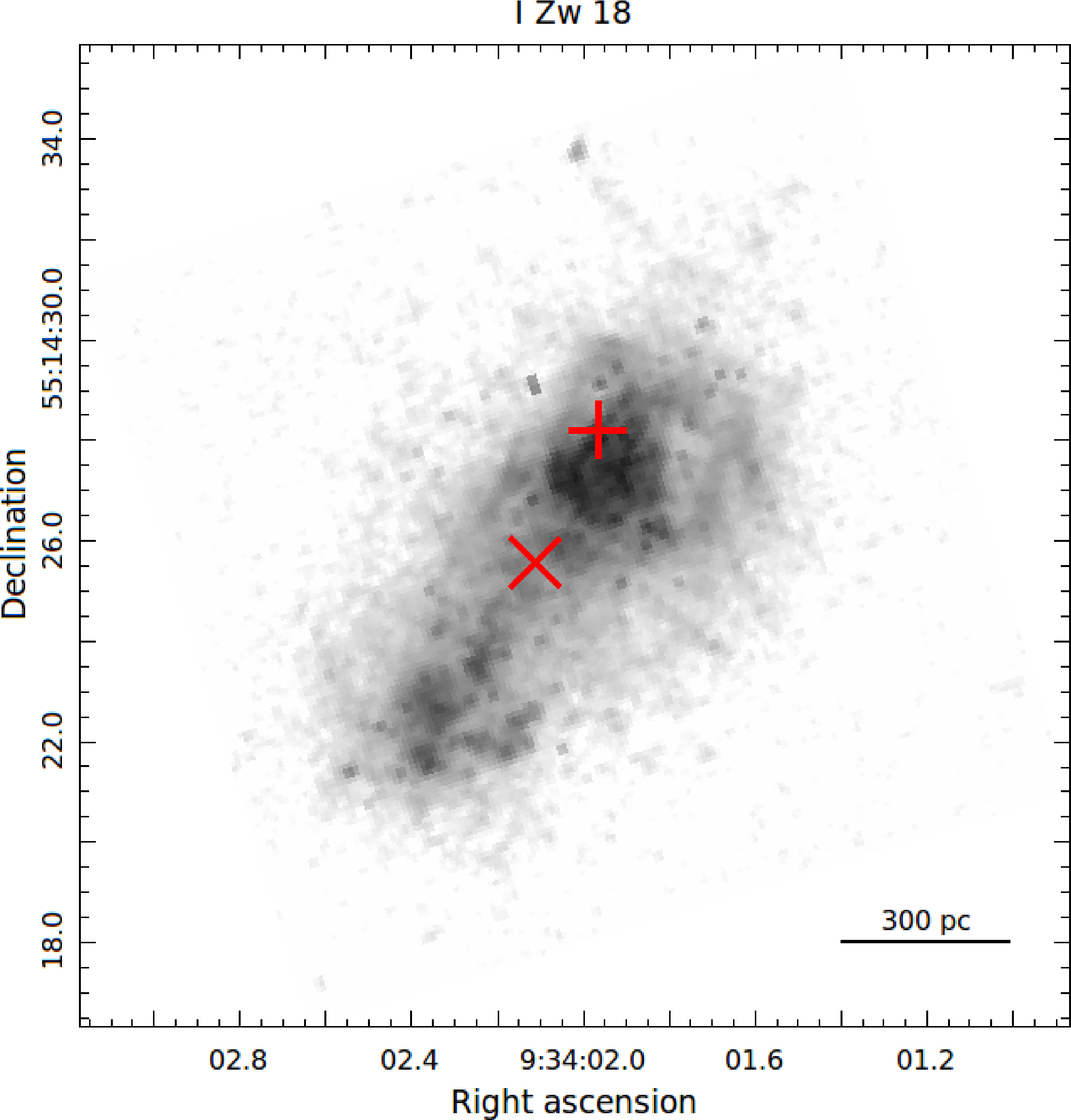}
\caption{HST images with X-ray source (red cross) and dynamical centers (red $\times$, \citep{lelli2012,simpson2011}). I~Zw~18 (right) has a luminous X-ray source within a dense star forming region of the galaxy. VII~Zw~403 (left) contains a X-ray source that is not near a dense star forming region and far from the dynamical center of the BCD. }\label{fig:hst_images}
\end{figure*}

The difference in the environments in which the X-ray sources of VII~Zw~403 and I~Zw~18 reside may provide an explanation for the differences in their spectral evolution. VII~Zw~403 and I~Zw~18 are dwarf galaxies with total stellar masses of $2 \times 10^7$~M$_\odot$ and $5.9\times 10^6$~M$_\odot$~\citep{zhao2013}, star formation rates of $1.4\times 10^{-2}$~M$_\odot$~yr$^{-1}$ and $7.0\times 10^{-2}$~M$_\odot$~yr$^{-1}$~\citep{brorby2014,prestwich2013}, and metallicities of $Z/Z_\odot = 0.062$ and $Z/Z_\odot = 0.019$~\citep{izotov2007}, respectively. In Figure~\ref{fig:hst_images}, one sees that the source in I~Zw~18 lies in a dense star-forming region, whereas the object in VII~Zw~403 is far from the dynamical center, away from dense regions. Intermediate mass black holes are more likely to form in young, dense clusters where dynamical friction causes the runaway merger of stars over a relatively short timescale, $t_\text{df} \leq 4$~Myr~\citep{portegies2004}. The spectra of these objects (if in a binary system) are expected to behave as scaled-up versions of the Galactic black hole binaries. The fact that I~Zw~18 exhibits this type of spectral state behavior and that it resides in a dense environment does not rule out the possibility that it is an intermediate mass black hole binary. However, for the source in VII~Zw~403, given its location and spectral shape, it seems unlikely that it could be an intermediate mass black hole. More observations that catch these objects in a broad range of luminosity states would be needed to conclusively determine their spectral state transition history and thus the nature of their primary compact objects. Recent results by \cite{bachetti2014} illuminate the need for more observations by showing that a known ULX in the nuclear region of M82, which contains many dense star clusters, is a super-Eddington accreting neutron star and not a massive black hole.

Most extragalactic ULXs that exhibit spectral curvature are located away from the galactic nucleus and removed from very dense regions~\citep{pakull2006,roberts2007,grise2008}. These ULXs, based on their location and spectra, are argued to be stellar mass black hole binaries emitting at super-Eddington luminosities, though the mechanism by which this happens is not completely understood. VII~Zw~403 observations are more consistent with these properties, suggesting it may be a super-Eddington stellar mass black hole binary. The transient behavior of VII~Zw~403 source is particularly interesting since most ``standard'' ULXs are nontransient and may be active for up to decades in some cases~\citep{kaaret2009,feng2011}. In this way, VII~Zw~403 is more like the Galactic BHBs, spending most of its time in the quiescent state. More observations of the low and transition states may provide a means to understand the connection between the ultraluminous state and the states commonly seen in Galactic black hole X-ray binaries.

\subsection{Implications}
Study of nearby HMXBs can provide insight into the formation of high mass X-ray binaries formed in the early universe. Observational and theoretical studies have shown that reduced metallicity enhances X-ray binary production and the total X-ray luminosity of a galaxy. The total luminosity, and hence the spectral shape of the emission from a population of HMXBs, is dominated by a relatively small population of sources in high luminosity states. To date, a total of eight low-metallicity BCDs have been found that contain ULXs~\citep{thuan2014} as defined by having a luminosity $> 10^{39}$~erg~s$^{-1}$ (Eddington luminosity of a 10 M$_\odot$ black hole). Only two of these, VII~Zw~403 and I~Zw~18, have had sufficient counts to allow for a spectral study and produced inconsistent results. Our discovery of this state transition of an X-ray source in a low-metallicity blue compact dwarf galaxy shows that the sources may undergo a transition to a spectral state with high-energy cutoffs at high luminosity, similar to those seen in other ULXs.

Other low-metallicity, star-forming dwarf galaxies, such as Holmberg~II~\citep{goad2006} and Holmberg~IX~\citep{stobbart2006,gladstone2009,walton2014}, contain known ULXs that exhibit spectra consistent with the ultraluminous state, as defined by \cite{gladstone2009}. The metallicities of these galaxies are low $(12+\log($O/H$) < 8.0)$, but do not meet the criteria for being XMPGs $(12+\log($O/H$) < 7.7)$, mostly occupied by BCDs. Low metallicity galaxies, such as Ho~II and Ho~IX, may be useful for studies of ULX spectral evolution with metallicity.

These results motivate further spectral studies of luminous HMXBs in nearby analogues of early universe galaxies. This will allow for better estimates of the spectral shape of HMXBs in the early universe, which affect the ionisation morphology during the Epoch of Reionization via thermal feedback, and the time at which reionisation was completed.

\section{ACKNOWLEDGEMENTS}
We thank the anonymous referee for helpful comments and suggestions.
This research has made use of data obtained from the Suzaku satellite, a collaborative mission between the space agencies of Japan (JAXA) and the USA (NASA).

\label{lastpage}

\end{document}